# MEASUREMENT AND CORRECTION OF CROSS-PLANE COUPLING IN TRANSPORT LINES*

M. Woodley, P. Emma, *SLAC*, Stanford, CA 94309, USA

## *Abstract*

In future linear colliders the luminosity will depend on maintaining the small emittance aspect ratio delivered by damping rings. Correction of cross-plane coupling can be important in preventing dilution of the beam emittance. In order to minimize the vertical emittance, especially for a flat beam, it is necessary to remove all cross-plane (*x*-*y*) correlations. This paper studies emittance measurement and correction for coupled beams in the presence of realistic measurement errors. The results of simulations show that reconstruction of the full 4×4 beam matrix can be misleading in the presence of errors. We suggest more robust tuning procedures for minimizing linear coupling.

## 1 INTRINSIC EMITTANCE

A four-dimensional (4D) symmetric beam matrix, $\sigma$, contains ten unique elements, four of which describe coupling. The *projected* (2D) beam emittances, $\varepsilon_x$ and $\varepsilon_y$, are defined as the square roots of the determinants of the *on*-diagonal 2×2 submatrices. If one or more of the elements of the *off*-diagonal submatrix is non-zero, the beam is *x*-*y* coupled. Diagonalization of the beam matrix yields the *intrinsic* beam emittances, $\varepsilon_1$ and $\varepsilon_2$.

$$\sigma = \begin{pmatrix} <x^2> & <xx'> & <xy> & <xy'> \\ <xx'> & <x'^2> & <x'y> & <x'y'> \\ <xy> & <x'y> & <y^2> & <yy'> \\ <xy'> & <x'y'> & <yy'> & <y'^2> \end{pmatrix}, \quad \bar{\sigma} = \bar{R}\sigma\bar{R}^T = \begin{pmatrix} \varepsilon_1 & 0 & 0 & 0 \\ 0 & \varepsilon_1 & 0 & 0 \\ 0 & 0 & \varepsilon_2 & 0 \\ 0 & 0 & 0 & \varepsilon_2 \end{pmatrix}$$

The coupling correction process involves measuring the ten elements of the beam matrix and finding a set of skew quadrupole strengths which block diagonalize the beam matrix, setting the projected emittances, for linear coupling, equal to the intrinsic emittances.

## 2 SKEW CORRECTION SECTION

The ideal skew correction section (SCS) contains four skew quadrupoles separated by appropriate betatron phase advance in each plane such that the skew quadrupoles are orthonormal (orthogonal and equally scaled). A simple realization of such a system is possible if the skew quadrupoles each correct just one of the four *x*-*y* beam correlations and if, in addition, the product $\beta_x\beta_y$ is equal at each of the skew quadrupoles. Figure 1 shows such a system for the 250 GeV NLC beam, followed by a 4D emittance measurement section (described below). Skew quadrupoles at locations 1-4 (indicated at top of figure by diamond symbols) are used to correct the <*xy*>, <*x′y′*>, <*x′y*>, and <*xy′*> beam correlations, respectively, at location 4. The horizontal and vertical betatron phase advances between the skew quadrupoles are also indicated on the figure. This scheme allows total correction of any arbitrary linearly coupled beam with correction range limited only by the available skew quadrupole strength.

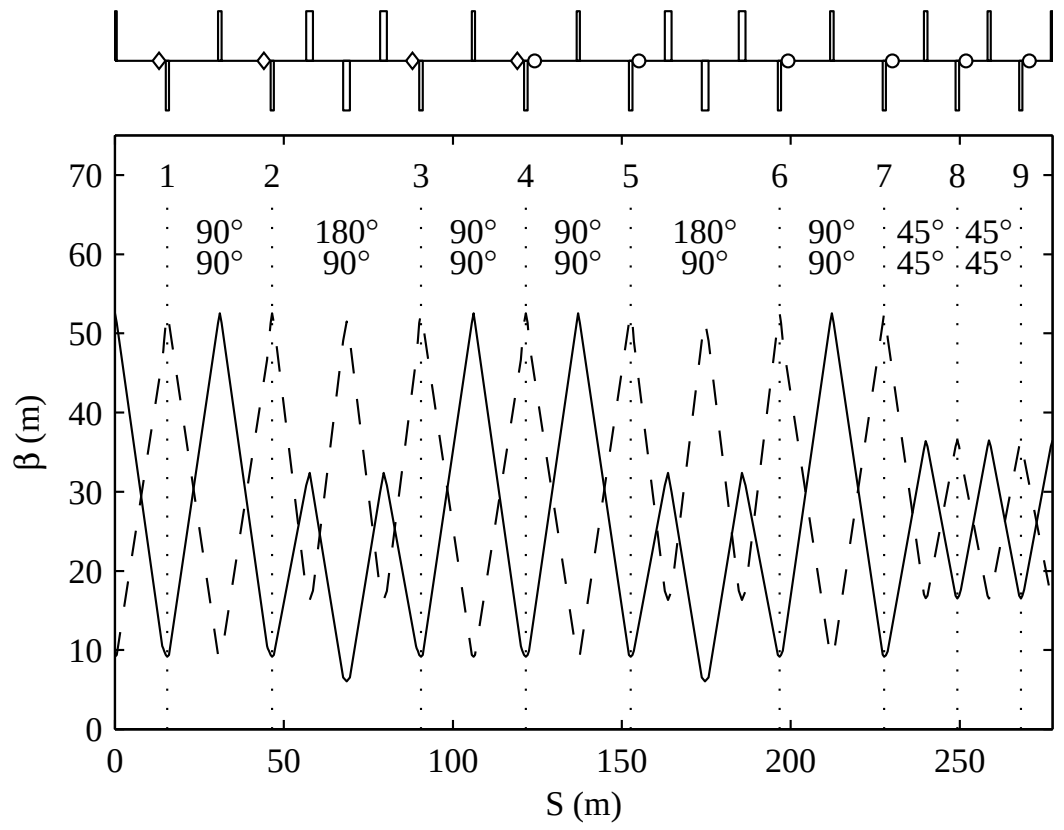

Figure 1: SCS (*S*=0-120 m) plus 4D emittance measurement section (*S*=120-270 m): $\beta_x$ (solid), $\beta_y$ (dash). Diamond symbols indicate skew quadrupoles; circles indicate wire scanners. The betatron phase advances between devices are shown in 2 rows above the plotted $\beta$–functions (*x* on top and *y* below).

## 3 4D EMITTANCE MEASUREMENT

The ideal 4D emittance measurement section contains six beam size measurement devices (e.g. wire scanners) separated by appropriate betatron phase advance in each plane such that the four *x*-*y* beam correlations may be measured independently. Figure 1 illustrates such a system. The wire scanners at locations 4-7 (circle symbols) are used to measure the <*xy*>, <*x′y′*>, <*x′y*>, and <*xy′*> beam correlations, respectively. Each wire scanner has three independent angle filaments — an *x*-wire, a *y*-wire, and an "off-axis", or *u*-wire whose optimal orientation is given by the inverse tangent of the uncoupled beam aspect ratio, $\sigma_y/\sigma_x$ [1]. At each of these wire scanners $\sigma_x$, $\sigma_y$, and $\sigma_{xy}$ are measured. An additional two wire scanners (locations 8 and 9 in Figure 1) are required to determine the remaining in-plane correlations of the beam. There are a total of 10 beam parameters to determine ($\varepsilon_{x,y}$, $\beta_{x,y}$, $\alpha_{x,y}$, and the four *x*-*y* correlations) and up to 18 beam size measurements, leaving 8 degrees of freedom in the analysis. The analysis consists of expressing the beam sizes at each wire in terms of the unknown beam parameters at the first wire, using the wire-to-wire R-matrices, and solving the linear system.

* Work supported by the U.S. Department of Energy under Contract DE-AC02-76SF00515.

Figures 2 and 3 each show the results of 5000 Monte Carlo simulations of the 4D analysis and intrinsic vertical emittance calculation using this setup. The input beam is the nominal NLC beam at 250 GeV ($\gamma\varepsilon_1$=3×10$^{-6}$ m, $\gamma\varepsilon_2$=3×10$^{-8}$ m). For these emittances, the ideal rms beam sizes at the wires range from 1.5-10 μm. In each simulation, the real beam size on each wire is given a gaussian distributed multiplicative random error of rms $f_{err}$

$$\sigma_{sim} = (1 + f_{err})\sigma_{ideal}$$

and the ensemble of simulated measurements is analyzed.

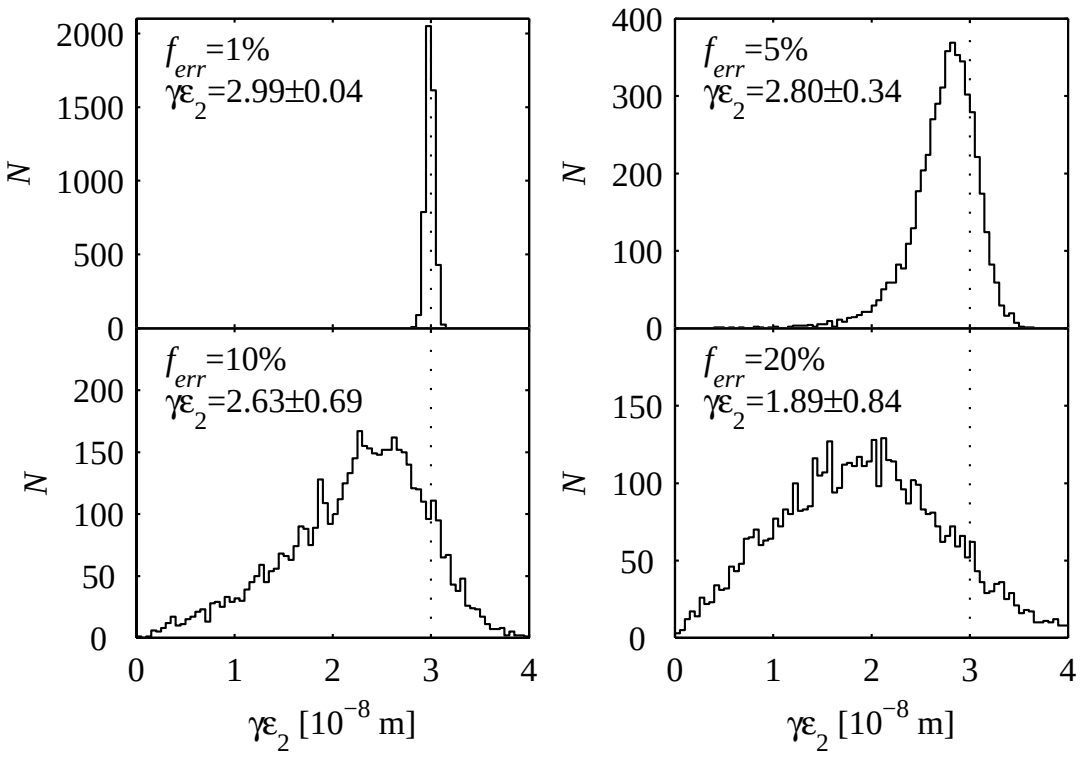

Figure 2: Results of simulations of 4D emittance measurement and reconstruction of $\gamma\varepsilon_2$ (coupled input beam). Vertical dotted lines show the actual value $\gamma\varepsilon_{20}$ used in the simulations.

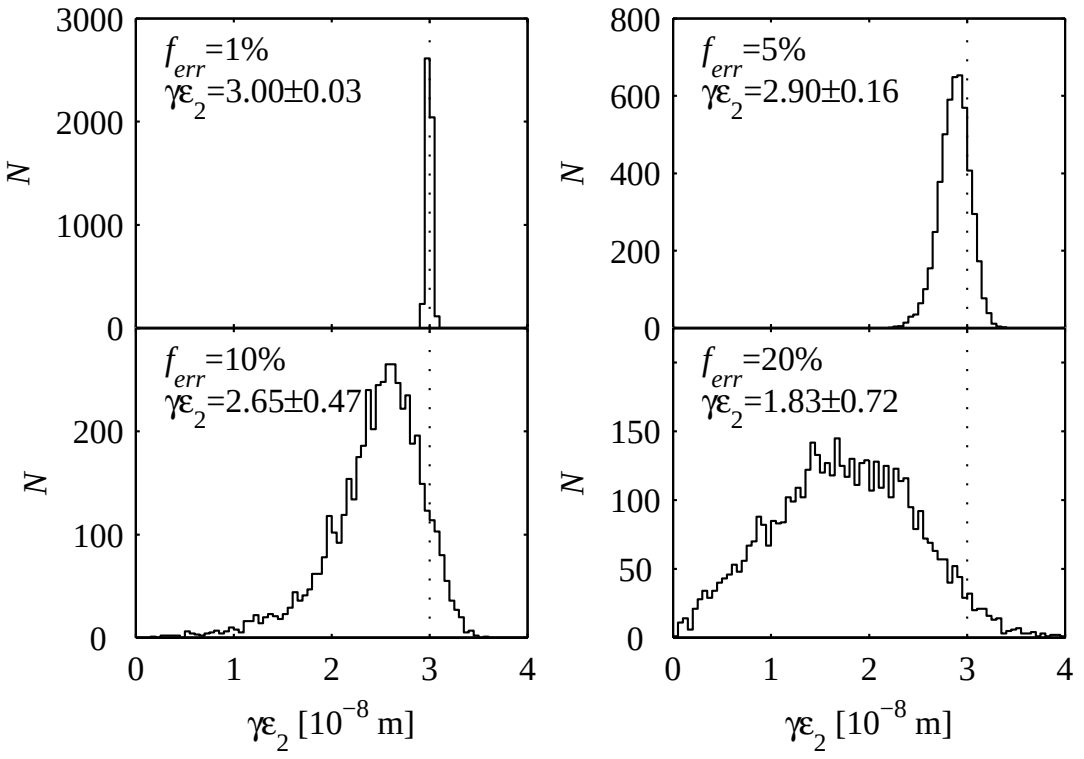

Figure 3: Results of simulations of 4D emittance measurement and reconstruction of $\gamma\varepsilon_2$ (uncoupled input beam).

Figure 2 shows the results for four values of $f_{err}$ when the simulated input beam is coupled ($\varepsilon_y/\varepsilon_2$ = 1.5), while Figure 3 shows the results for an uncoupled input beam ($\varepsilon_y/\varepsilon_2$ = 1). Figures 2 and 3 show that when the beam size measurement errors are more than a few percent, the measurements become imprecise, and more importantly, the most probable computed value for the intrinsic vertical emittance becomes erroneously small. This bias may lead one to attempt to correct the implied coupling, which will actually introduce coupling rather than correct it. An additional problem, in the presence of errors, is that the 4D analysis can generate solutions for which the beam matrix is nonpositive, yielding imaginary emittances. As $f_{err}$ becomes larger, the fraction of simulations which yield nonpositive beam matrices, the 'rejection fraction', increases to the point where 3 out of 4 measurements yield non-physical results when $f_{err}$ reaches 20 %. Table 1 summarizes the results of the 4D measurement simulations for a coupled input beam; Table 2 summarizes the results for an uncoupled input beam. In each case, the most probable relative value of $\varepsilon_2/\varepsilon_{20}$ is given, along with the statistical rms width of the distribution (where $\varepsilon_{20}$ is the 'real' intrinsic emittance used in the simulations).

Table 1: 4D Simulation Results (coupled beam).

| $f_{err}$ | $\varepsilon_2/\varepsilon_{20}$ | *rejection fraction* |
|---|---|---|
| 1 % | 1.00 ± 0.01 | <0.1 % |
| 5 % | 0.93 ± 0.10 | 0.2 % |
| 10 % | 0.88 ± 0.23 | 22 % |
| 20 % | 0.63 ± 0.28 | 78 % |

Table 2: 4D Simulation Results (uncoupled beam).

| $f_{err}$ | $\varepsilon_2/\varepsilon_{20}$ | *rejection fraction* |
|---|---|---|
| 1 % | 1.00 ± 0.01 | <0.1 % |
| 5 % | 0.97 ± 0.05 | <0.1 % |
| 10 % | 0.88 ± 0.16 | 1.9 % |
| 20 % | 0.61 ± 0.24 | 59 % |

# 4 2D EMITTANCE MEASUREMENT

An optimized 2D emittance measurement section contains four wire scanners separated by 45° of betatron phase advance in both planes. Figure 4 shows such a system preceded by an SCS. Each wire scanner has two independent angle filaments—an *x*-wire and a *y*-wire. At each wire scanner $\sigma_x$ and $\sigma_y$ are measured. There are a total of three beam parameters to determine ($\varepsilon$, $\beta$ and $\alpha$) and four beam size measurements in each plane, leaving one degree of freedom in the analysis for each plane.

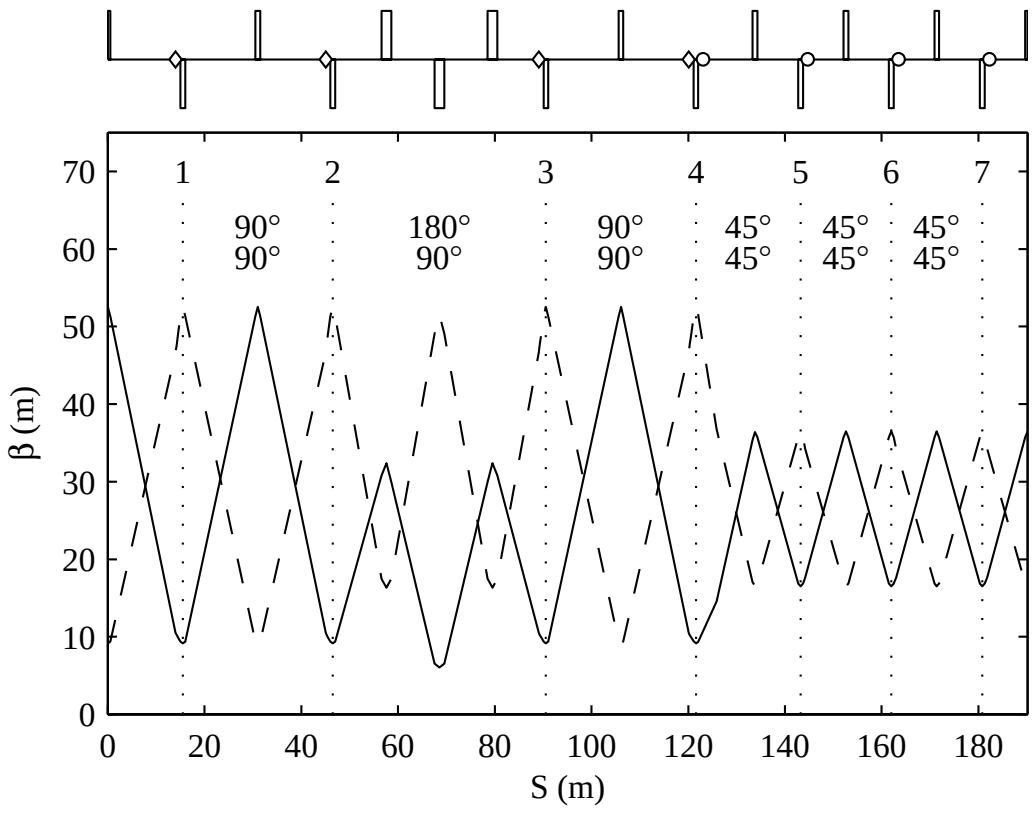

Figure 4: SCS (*S*=0-120 m) plus 2D emittance measurement section (*S*=120-190 m): $\beta_x$ (solid), $\beta_y$ (dash).

Figures 5 and 6 each show simulations of the 2D analysis and projected vertical emittance calculation using this setup. Figure 5 is for a coupled input beam, while Figure 6 is for an uncoupled input beam.

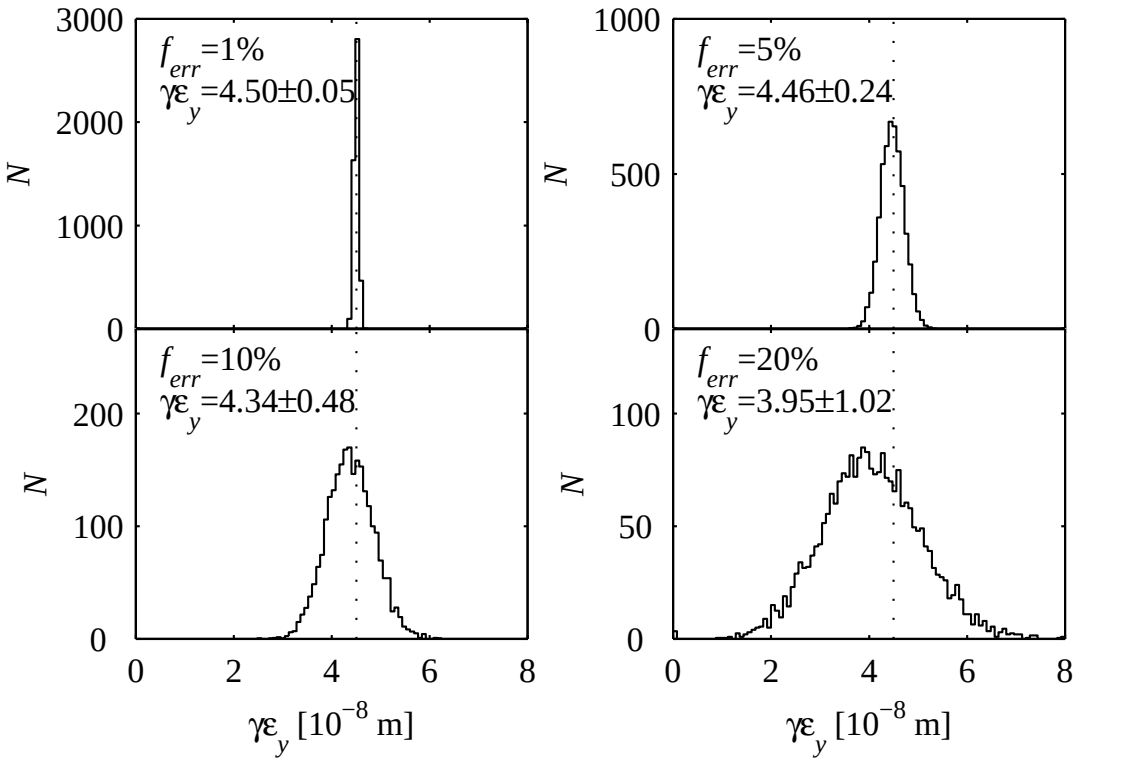

Figure 5: Results of simulations of 2D emittance measurement and reconstruction of $\gamma\varepsilon_y$ (coupled input beam). Vertical dotted lines show the actual value $\gamma\varepsilon_{y0}$ used in the simulations.

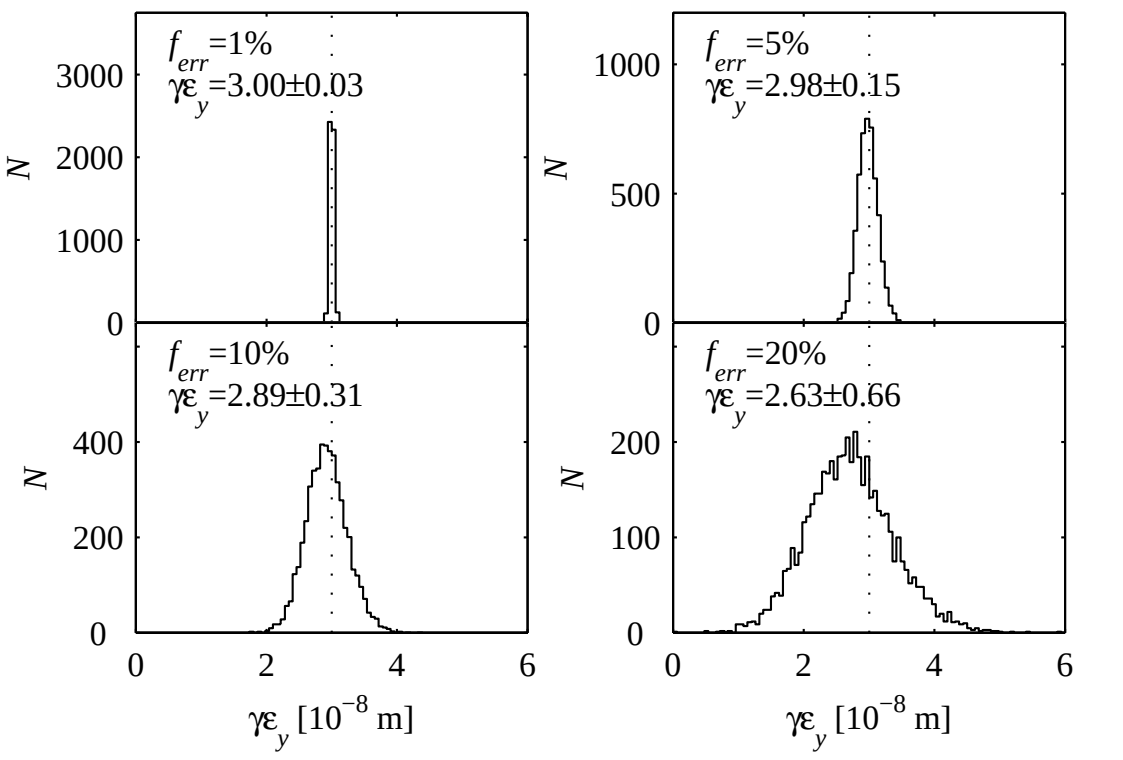

Figure 6: Results of simulations of 2D emittance measurement and reconstruction of $\gamma\varepsilon_y$ (uncoupled input beam).

These figures show that the 2D projected emittance measurement is far less sensitive to beam size measurement errors than the 4D intrinsic emittance measurement In addition, the 2D analysis does not generate non-physical solutions. Table 3 summarizes the 2D measurement simulations for a coupled input beam; Table 4 summarizes the results for an uncoupled input beam ($\varepsilon_{y0}$ is the 'real' projected emittance).

Table 3: 2D Simulation Results (coupled beam).

| $f_{err}$ | $\varepsilon_y / \varepsilon_{y0}$ | *rejection fraction* |
|---|---|---|
| 1 % | 1.00 ± 0.01 | 0 |
| 5 % | 0.99 ± 0.05 | 0 |
| 10 % | 0.96 ± 0.11 | 0 |
| 20 % | 0.88 ± 0.23 | 0 |

Table 4: 2D Simulation Results (uncoupled beam).

| $f_{err}$ | $\varepsilon_y / \varepsilon_{y0}$ | *rejection fraction* |
|---|---|---|
| 1 % | 1.00 ± 0.01 | 0 |
| 5 % | 0.99 ± 0.05 | 0 |
| 10 % | 0.96 ± 0.10 | 0 |
| 20 % | 0.88 ± 0.22 | 0 |

# 5 COUPLING CORRECTION

Given the unreliability of the 4D emittance measurement, we propose, for the NLC, the coupling correction and 2D emittance measurement system shown in Figure 4. Coupling correction will be achieved by sequentially minimizing the measured projected vertical emittance with each of the four orthonormal skew quadrupoles. Figure 7 shows the Monte Carlo simulation of this process, assuming a coupled input beam ($\varepsilon_y/\varepsilon_2 > 3$) and 10% beam size measurement errors. Because the optics of the SCS has been designed to make the skew quadrupoles orthonormal, a single pass through the set is sufficient to bring the projected vertical emittance down to its intrinsic value to within measurement errors.

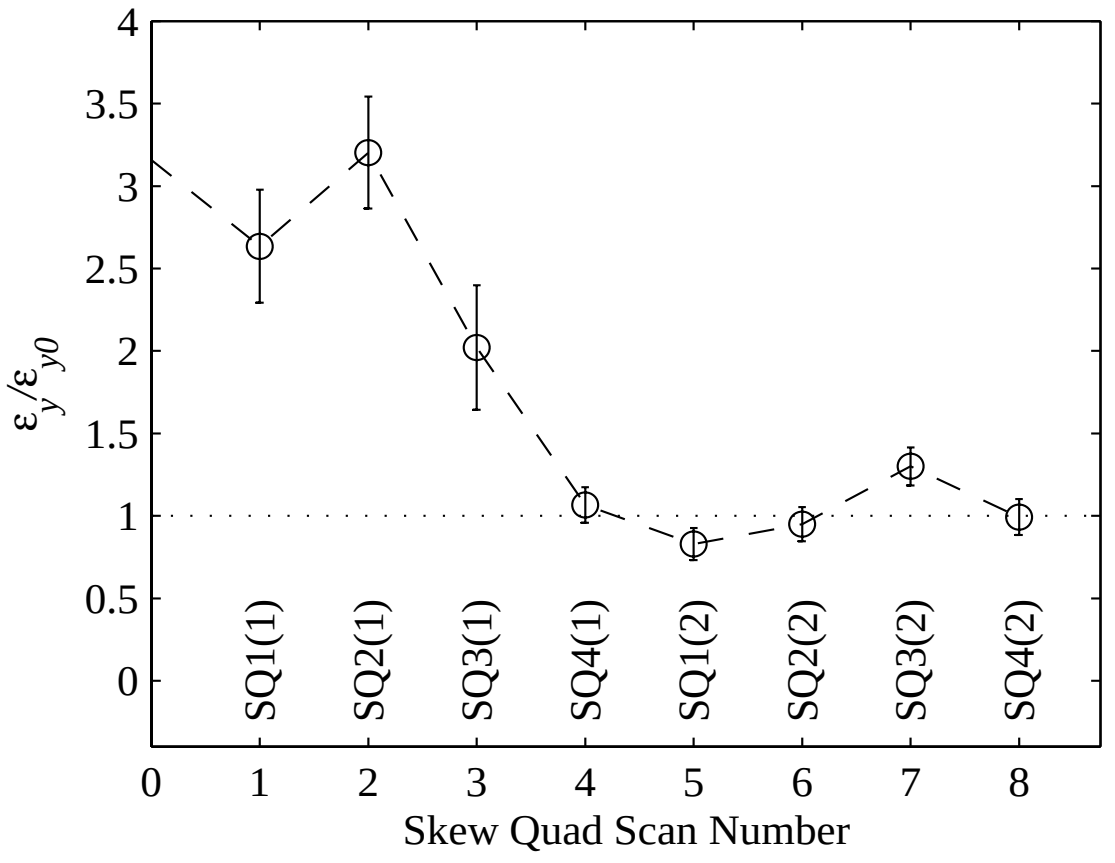

Figure 7: Results of simulations of two full iterations of coupling correction. Each circle gives the minimized value of $\varepsilon_y/\varepsilon_2$ after scanning the indicated skew quad.

Alternatively, the system shown in Figure 1 can be used to remove the coupling more directly. Each skew quadrupole can be used to remove the measured <*xy*> correlation at its associated wire scanner (skew quadrupoles 1-4 correct <*xy*> at wire scanners 4-7, respectively).

# 6 CONCLUSIONS

Although it may seem that the 4D emittance measurement is the most direct way to compute skew corrections for a coupled beam, simulations show that realistic beam size measurement errors degrade the analysis to the point where it becomes counter-productive. The 2D emittance measurement is far more reliable, and when combined with an orthonormal skew correction system, provides the most robust method for correcting linear betatron coupling.